\documentclass[]{aa} % for printer version
\usepackage{natbib}
\bibpunct{(}{)}{;}{a}{}{,} % to follow the A&A style
\usepackage{pbox}
\usepackage{tablefootnote}
\usepackage{url}
\usepackage{amssymb}

\usepackage{graphicx}
\usepackage{txfonts}
\begin{document}

   \title{Search for an exosphere in sodium and calcium in the transmission spectrum of exoplanet 55 Cancri e \thanks{Based on observations collected at the European Organisation for Astronomical Research in the Southern Hemisphere under ESO programmes 092.C-0178 \& 288.C-5010 and the Telescopio Nazionale Galileo under programme CAT13B\_33.}} 

   %\subtitle{}

   \author{A. R. Ridden-Harper
          \inst{1}
          \and
          I. A. G. Snellen \inst{1}
          \and
          C. U. Keller \inst{1}
          \and R. J. de Kok \inst{1} 
          \and E. Di Gloria \inst{1} 
          \and H. J. Hoeijmakers \inst{1} 
          \and M. Brogi \inst{2}
          \and M. Fridlund \inst{1,3,4}
          \and B. L. A. Vermeersen \inst{5}
          \and W. van Westrenen \inst{6}}

   \institute{Leiden Observatory, Leiden University, Niels Bohrweg 2, 2333 CA Leiden, The Netherlands
                  \\
              \email{arh@strw.leidenuniv.nl}              
             \and 
             Center for Astrophysics and Space Astronomy, University of Colorado at Boulder, Boulder, CO 80309, USA             
              \and
             Max Planck Institute for Astronomy, K\"onigstuhl 17, 69117 Heidelberg, Germany             
             \and 
             Department of Earth and Space Sciences, Chalmers University of Technology, Onsala Space Observatory, 439 92 Onsala, Sweden             
             \and             
             Delft University of Technology, The Netherlands
             \and             
             Department of Earth Sciences, Vrije Universiteit Amsterdam, De Boelelaan 1085, 1081 HV Amsterdam, Netherlands}

   \date{}

% \abstract{}{}{}{}{} 
% 5 {} token are mandatory
 
  \abstract
   {The atmospheric and surface characterization of rocky planets is a key goal of exoplanet science. Unfortunately, the measurements required for this are generally out of reach of present-day instrumentation. However, the planet Mercury in our own solar system exhibits a large exosphere composed of atomic species that have been ejected from the planetary surface by the process of sputtering. Since the hottest rocky exoplanets known so far are more than an order of magnitude closer to their parent star than Mercury is to the Sun, the sputtering process and the resulting exospheres could be orders of magnitude larger and potentially detectable using transmission spectroscopy, indirectly probing their surface compositions.}
   {The aim of this work is to search for an absorption signal from exospheric sodium (Na) and singly ionized calcium (Ca$^+$) in the optical transmission spectrum of the hot rocky super-Earth 55 Cancri e. Although the current best-fitting models to the planet mass and radius require a possible atmospheric component, uncertainties in the radius exist, making it possible that 55 Cancri e could be a hot rocky planet without an atmosphere.}
   {High resolution (R$\sim$110000) time-series spectra of five transits of 55 Cancri e, obtained with three different telescopes (UVES/VLT, HARPS/ESO 3.6m \& HARPS-N/TNG) were analysed. Targeting the sodium D lines and the calcium H and K lines, the potential planet exospheric signal was filtered out from the much stronger stellar and telluric signals, making use of the change of the radial component of the orbital velocity of the planet over the transit from $-$57 to +57 km sec$^{-1}$.}
   {Combining all five transit data sets, we detect a signal potentially associated with sodium in the planet exosphere at a statistical significance level of 3$\sigma$. Combining the four HARPS transits that cover the calcium H and K lines, we also find a potential signal from ionized calcium (4.1 $\sigma$). Interestingly, this latter signal originates from just one of the transit measurements - with a 4.9$\sigma$ detection at this epoch. Unfortunately, due to the low significance of the measured sodium signal and the potentially variable Ca$^+$ signal, we estimate the  p-values of these signals to be too high (corresponding to $<$4$\sigma$) to claim unambiguous exospheric detections. 
 By comparing the observed signals with artificial signals injected early in the analysis, the absorption by Na and Ca$^+$ are estimated to be at a level of $\sim 2.3\times 10^{-3}$ and $\sim 7.0\times 10^{-2}$ respectively, relative to the stellar spectrum.}
   {If confirmed, the $3\sigma$ signal would correspond to an optically thick sodium exosphere with a radius of 5 $ \mathrm{R_\oplus}$, which is comparable to the Roche lobe radius of the planet.  The 4.9$\sigma$ detection of Ca$^+$ in a single HARPS data set would correspond to an optically thick Ca$^+$ exosphere approximately five times larger than the Roche lobe radius.  If this were a real detection, it would imply that the exosphere exhibits extreme variability. Although no formal detection has been made, we advocate that probing the exospheres of hot super-Earths in this way has great potential, also knowing that Mercury's exosphere varies significantly over time. It may be a fast route towards the first characterization of the surface properties of this enigmatic class of planets.}  


   \maketitle
%
%________________________________________________________________

\section{Introduction}

Transit and radial velocity surveys have revealed a new class of rocky planets orbiting their parent stars at very short distances (0.014 - 0.017 au\footnote{0.014 is the semi-major axis of GJ 1214b, identified from exoplanets.org as the shortest semi-major axis with filter MINSI[mjupiter] < 10M$_\oplus$ and 0.017 comes from the definition of an ultra-short period planet (USP) from \cite{Demory2015} of P < 0.75 days.}). Their evolutionary history is unknown. They may be rocky planets formed at significantly larger distances that subsequently migrated inwards, or could originally be gas-rich planets which lost their gaseous envelopes during migration through tidal heating and/or direct stellar irradiation \citep{Raymond2008}. Insights into the composition of these hot rocky planets would help to distinguish between the different scenarios. Although the first secondary eclipse measurements have been presented in the literature \citep{Demory2012}, showing them to be indeed very hot, with observed surface or atmospheric temperatures of 1300 - 3000K, detailed observations that could reveal atmospheric or surface compositions are beyond the reach of current instrumentation. 

One physical process that could reveal information of a planet's surface composition, potentially also with current instruments, is that of sputtering. Atomic species are ejected from the planet surface by the intense stellar wind of charged particles, creating an extended exosphere around the planet. This process is well known from planet Mercury in our own solar system.  It has an exosphere composed of atomic species including sodium (Na), calcium (Ca) and magnesium (Mg), which are thought to be the results of sputtering, thermal vaporisation, photon-stimulated desorption, and meteoroid impact vaporisation. Since the discovery of an emission spectrum of sodium in the exosphere of Mercury by \cite{Potter1985}, it has been subsequently detected many times \citep[see][for a review]{Killen2007} in emission, and less commonly in absorption during the transit of Mercury in front of the Sun \citep{Potter2013}.  These decades of observations have revealed that sodium in the exosphere of Mercury shows a great deal of spatial and temporal variability.  Rapid variations at a 50\% level on timescales of a day in the ion-sputtering component of the sodium in Mercury's exosphere due to variability in the magnetosphere have been observed, as well as latitudinal and/or longitudinal variations \citep{Potter1990,Killen2007}. In addition, long-term variations on timescales of months to years in photon-stimulated desorption \citep{Lammer2003,Killen2007} and radiation pressure acceleration 
\citep{Smyth1995,Killen2007} have been observed, as well as variations in meteoritic vaporisation \citep{Morgan1988,Killen2007}. 

\cite{Mura2011} argued for the first time that such exospheres resulting from the sputtering process may be observable for hot rocky exoplanets. Their simulations for CoRoT-7b suggest that it may have a high escape rate of species such as Na, Ca$^+$, and Mg$^+$ which likely form a tail tens of planetary radii long. \cite{Guenther2011} observed a transit of CoRoT-7b with the UVES instrument on the VLT with a focus on Na, Ca, and Ca$^+$. While \cite{Guenther2011} express their derived upper limits in units of stellar luminosity ($2-6\times10^{-6}L_{*}$), these limits appear to correspond to an absorption level on the order of approximately 3$\times$10$^{-3}$ smeared out over a 55 km sec$^{-1}$ velocity bin due to the change in the radial component of the orbital velocity of the planet during their long exposures. In this paper we target the exoplanet 55 Cancri e, whose host star has an apparent magnitude of $V=5.95$, 200 times brighter than CoRoT-7.

In addition, \cite{Schaefer2009} argue that a tidally locked hot rocky super-Earth could have a magma ocean that releases vapours to produce a silicate based atmosphere.  Their models show that Na is likely the most abundant constituent of such an atmosphere, which they believe could form a large cloud of Na through interaction with the stellar wind. 

Considerable progress regarding the detection and study of exospheres of hot gas giant exoplanets has already been made.  Hydrogen exospheres extending beyond the Roche lobe have been repeatedly detected around HD 209458b \citep{Vidal-Madjar2003,Vidal-Madjar2004} and HD 189733b \citep{Lecavelier_des_Etangs2010}, where the hydrogen signal from HD 189733b has been claimed to show temporal variation \citep{Lecavelier_des_Etangs2012}. Heavier atoms and ions have been detected in the exosphere of HD 209458b, including C$^+$ \citep{Vidal-Madjar2004,Linsky2010} and, more tentatively, O \citep{Vidal-Madjar2004}, Mg \citep{Vidal-Madjar2013} and Si$^{2+}$ \citep{Linsky2010}. Exospheric studies have recently also been extended to smaller planets with the detection of hydrogen around the warm Neptune GJ 436b \citep{Ehrenreich2015}. We note that no hydrogen exosphere was detected around 55 Cancri e \citep{Ehrenreich2012}, which is the object of this study.

The hot, rocky super-Earth type planet, 55 Cancri e (or $\mathbf{\rho^1}$ Cancri e, 55 Cnc e) orbits a bright (V=5.95) 0.905 $M_\sun$ star. It has a very short orbital period of 17.7 hours (see Table \ref{table:55Cncparameters} for uncertainties), a radius of $2.173$ $R_\oplus$ \citep{Gillon2012}, a mass of $8.09$ $M_\oplus$, and an inferred average density of $5.51$ $\mathrm{g cm^{-3}}$ \citep{Nelson2014}.  Transits of 55 Cnc e have been detected with broadband photometry from space in the visible \citep{Winn2011} and infra-red \citep{Demory2011}, and recently also from the ground \citep{deMooij2014}.

There is significant debate in the literature about the chemical composition and interior structure of 55 Cnc e. Using the internal structure model by \cite{Valencia2006,Valencia2010}, \cite{Gillon2012} argue that 55 Cnc e is likely a rocky, oxygen-rich planet composed of silicates with a gaseous envelope consisting of either a mixture of hydrogen and helium of approximately 0.1\% by mass or a water atmosphere of approximately 20\% by mass. However, because such a low mass H-He atmosphere would escape over a timescale of millions of years, while a water-vapour atmosphere could survive over billions of years, the water-vapour atmosphere interpretation is favoured, where the water-vapour is in a super-critical form due to its high temperature.   Furthermore, \cite{Ehrenreich2012} found that 55 Cnc e lacks a H exosphere which could be the result of complete H loss from the atmosphere in the past.  In contrast,  \cite{Madhusudhan2012} claim that if 55 Cnc e were to be a carbon-rich planet, a different structure is possible where Fe, C (in the form of graphite and diamond), SiC, and silicates of a wide range of mass fractions could explain its density without the need for a gaseous envelope. While the C/O ratio of 55 Cnc was previously thought to be $>$1, a subsequent analysis by \cite{Teske2013} found that it more likely has a C/O ratio of 0.8.  This value is lower than the value adopted by \cite{Madhusudhan2012} of 1.12$\pm$0.19; however, it still corresponds to the predicted minimum value of 0.8 necessary for the formation of a carbon-rich condensate under the assumption of equilibrium \citep{Larimer1975}.

Furthermore, \cite{Demory2015} report a 4$\sigma$ detection of variability in the day-side thermal emission of 55 Cnc e, with the emissions varying by a factor of 3.7 between 2012 to 2013.  They also tentatively suggest variations in the transit depth and calculate the planetary radii to range from 1.75$\pm$0.13 $R_\oplus$ to 2.25$\pm$0.17 $R_\oplus$ with a mean value of 1.92$\pm$0.08 $R_\oplus$, which is approximately 2$\sigma$ smaller than the value published by \cite{Gillon2012} of 2.17$\pm$0.10$R_\oplus$  based on Spitzer+MOST data. We believe that this smaller radius implies that the need for a significant atmosphere to explain the planet's radius is significantly reduced.

%*******************************************************************************
%*******************************************************************************
%*******************************************************************************

If 55 Cnc e does not have an atmosphere, its surface would be directly exposed to stellar radiation, making it analogous to Mercury. It is likely that the processes which produce the exosphere of Mercury would be much more pronounced on 55 Cnc e because it receives a bolometric flux from its host star that is approximately 500 times greater than Mercury receives from the Sun. This corresponds to an equilibrium temperature of 55 Cnc e of almost 2000 K.  \cite{Demory2015} claim to have detected brightness temperatures which vary from 1300 K to 3000 K; however, the mechanism which causes this variability is not understood.  

\cite{Demory2016} report the observation of a complete phase curve of 55 Cnc e in the 4.5$\mu$m channel of the Spitzer Space Telescope Infrared Array Camera which allowed them to construct a longitudinal thermal brightness map due to 55 Cnc e being tidally locked to its host star.  This map revealed that 55 Cnc e has a strong day-night temperature contrast with temperatures of 2700 K and 1380 K on the day and night sides respectively.  Furthermore, they found that the day side exhibits highly asymmetric thermal emissions, with a hot spot located 41 degrees east of the substellar point.  These observations were interpreted as being either due to an atmosphere with heat recirculation confined to the day side only, or a planet without an atmosphere with low-viscosity magma flows on the surface.  Atmospheric escape rate arguments indicate that it is unlikely that 55 Cnc e has a thick atmosphere, so the magma ocean interpretation is favoured.

In this paper, we aim to search for an absorption signal from exospheric sodium (Na) and singly ionized calcium (Ca$^+$) in the optical transmission spectrum of the hot rocky super-Earth 55 Cnc e.
This paper is structured as follows. Section \ref{sec:obdata} describes the data and Section \ref{sec:analysis} explains the methods used in this analysis. Section \ref{sec:results} presents the results, and Section \ref{sec:discussion_and_conclusion} discusses the interpretation of the results and concludes.   

\begin{table}[]
\centering
\caption{Properties of 55 Cancri e} 
\label{table:55Cncparameters}
\begin{tabular}{@{}l@{}c@{}c@{}}
\hline \hline 
Parameter     &  Value  &  Source \\ \hline

\multicolumn{2}{l}{Stellar properties}\\
\\
Distance (pc) & 12.34 $\pm$ 0.11 & \cite{vanLeeuwen2007} \\
Radius (R$_\sun$) & 0.943$\pm$0.010 & \cite{vonBraun2011} \\
Luminosity (L$_\sun$) & 0.582 $\pm$ 0.014 & \cite{vonBraun2011} \\
T$_{EFF}$ (K) & 5196$\pm$24 & \cite{vonBraun2011} \\
Mass (M$_\sun$) & 0.905$\pm$0.015 & \cite{vonBraun2011}\\
log $g$  &   4.45$\pm$0.01  &  \cite{vonBraun2011}  \\
Radial velocity & 27.58 $\pm$ 0.07 & \cite{Nidever2002} \\
\\\hline
\multicolumn{2}{l}{Planet properties}\\
\\
Period (days) & $0.7365449 \pm 0.000005$ & \cite{Gillon2012} \\
Orb. radius (AU) & 0.0154$\pm$0.0001 & * \\
Radius $(R_\oplus)$ & $2.173 \pm 0.098$ & \cite{Gillon2012} \\
Mass $(M_\oplus)$ & $8.09\pm0.26$ &  \cite{Nelson2014} \\ 
Density $(\mathrm{g cm^{-3}})$ & $5.51\pm^{1.32}_{1.00}$ & \cite{Nelson2014} \\ \\ \hline
\multicolumn{2}{l}{$^*$Calculated using Kepler's third law.}\\
\end{tabular}
\end{table}

%__________________________________________________________________

\section{Observational data \label{sec:obdata}}

High-dispersion spectral time series of five transits of 55 Cnc e taken with three different telescopes were used for our analysis. The five data sets each cover one transit including observations just before and after the transit. 
We observed one transit using the Ultraviolet and Visual Echelle Spectrograph (UVES) \citep{Dekker2000} installed on the Nasmyth B focus of the Very Large Telescope (VLT) at the Paranal Observatory. Furthermore, we retrieved additional data sets from observatory archives. Two of these were observed with the High Accuracy Radial Velocity Planet Searcher (HARPS) \citep{Mayor2003} located at the ESO 3.6m Telescope at the La Silla Observatory, and two from its northern-hemisphere copy - HARPS-N \citep{Cosentino2012} located at the 3.6m Telescopio Nazionale Galileo at the Roque de los Muchachos Observatory.  An overview of all observations is shown in Table \ref{table:timing}. 
 
\subsection{UVES data}

138 UVES spectra were obtained of 55 Cnc.  The transit timing, dates, exposure times, observational cadence and phase coverage are presented in Table \ref{table:timing}.  The observations were made using the red arm of UVES, utilizing grating CD\#3 with a central wavelength of 580.0 nm, resulting in a wavelength range of $4726.5 - 6835.1$ $\AA$. A resolving power of $R\approx 110000$ was achieved using a slit width of 0.3$''$ and image slicer \#3 to minimize the slit losses.
Using no charge-coupled device (CCD) binning, the sampling is two pixels per spectral element \citep{D'Odorico2000}.

Unfortunately, cirrus clouds were present during our observations which considerably decreased the signal-to-noise ratio (S/N) in the spectra, ranging from S/N=180 during relatively good spells down to S/N=60 per pixel.

\subsection{HARPS data}
The HARPS data used for our analysis cover two transits and were originally taken for ESO programme 288.C-5010 (PI: A. Triaud) which was used by \cite{Lopez-Morales2014} to investigate the Rossiter-Mclaughlin effect in 55 Cnc e. We retrieved the pipeline-reduced data from the ESO Science Archive Facility\footnote{\url{http://archive.eso.org/cms.html}}. 

HARPS has a resolving power of R$\approx$115000 and a wavelength range of $\mathrm{3800 - 6910}$ $\AA$. It is enclosed in a vacuum vessel, pressure and temperature controlled to a precision of $\pm$0.01 mbar and $\pm$0.01 K respectively, resulting in a wavelength precision of $\lesssim$ 0.5 m s$^{-1}$ night$^{-1}$ \citep{Bonfils2013}. It has two fibres which feed the spectrograph with light from the telescope and calibration lamp.  The fibre aperture on the sky is 1$''$.  The CCD has a pixel size of 15 $\mathrm{\mu m}$ and a sampling of 3.2 pixels per spectral element \citep{Mayor2003}.  The transit timing, dates, exposure times, observational cadence, and phase coverage are presented in Table \ref{table:timing}.

\subsection{HARPS-N data}
The HARPS-N observations also cover two transits, and were originally taken in TNG Observing programme CAT13B\_33 (PI: F. Rodler), also to investigate the Rossiter McLaughlin effect by the same team \citep{Lopez-Morales2014}. The pipeline-reduced data was retrieved by us from the TNG data archive \footnote{\url{http://ia2.oats.inaf.it/archives/tng}}.  

HARPS-N is a copy of HARPS so its properties are all identical or very similar to HARPS at ESO. It has a slightly different wavelength range of $3830 - 6900$ $\AA$ and a sampling of 3.3 pixels per FWHM.  It also has a greater temperature stability than HARPS of 0.001 K, giving a short-term precision of 0.3 m s$^{-1}$ and a global long-term precision of better than 0.6 m$^{-1}$ \footnote{\url{http://www.tng.iac.es/instruments/harps/}}. The fibre aperture on sky and spectral resolution are identical to those of HARPS.  The transit timing, dates, exposure times, observational cadence, and phase coverage are presented in Table \ref{table:timing}. 

We note that an additional five publicly available data sets\footnote{TNG programme IDs: OPT12B\_13, OPT13B\_30, OPT14A\_34} of the transit of 55 Cnc e were obtained with HARPS-N by \cite{Bourrier2014} to investigate the Rossiter-McLaughlin effect.  The individual spectra of these data sets have exposure times of 360 seconds, which is twice the average exposure time of the data used in this study.  Therefore, we chose to not use these data sets because due to the very rapid change in the radial component of the orbital velocity of the planet (114 km s$^{-1}$ over the transit), any planet signature would be smeared out by ten pixels, significantly decreasing its peak.

%______________________________________________________________

\section{Data analysis \label{sec:analysis}}

In our analysis we concentrate on the H and K lines of ionized calcium (at 3968.47 \AA\ and 3933.66 \AA\ respectively) and the two sodium D lines (5889.95 \AA\  and 5895.92 \AA). While the sodium lines are covered by all data sets, the ionized calcium lines are only present in the HARPS and HARPS-N data and not in the UVES data. 

\subsection{Processing of UVES spectra \label{sec:UVESprocessing}}

The observed spectra are dominated by stellar and possible telluric absorption lines which are orders of magnitude stronger than the expected planet features. Since the stellar and telluric absorption lines are quasi-fixed in wavelength (the stellar lines change in radial velocity by approximately 1.4 m sec$^{-1}$ during the four hour observations \citep{McArthur2004}) and the radial component of the orbital velocity of the planet changes by tens of km sec$^{-1}$ during the transit, the change in the Doppler shift of the planet lines can be used to separate the planet signal from that of the star and the Earth's atmosphere. The procedure we used to carry out this processing is very similar to that used in previous work \citep[eg.][]{Snellen2010,Hoeijmakers2015} and the individual steps are summarised below.  

\begin{enumerate}

\item Extraction of wavelength calibrated 1D spectra. The UVES data were reduced using the standard ESO UVES reduction pipeline \citep{Ballester2000} which was executed with Gasgano\footnote{\url{https://www.eso.org/sci/software/gasgano.html}} and EsoRex \footnote{\url{http://www.eso.org/sci/software/cpl/esorex.html}}.  The pipeline produced a one-dimensional wavelength calibrated spectrum for each order for each exposure.  

\item Normalization of the spectra to a common flux level. Variation in instrumental throughput (for example, due to slit losses) and atmospheric absorption result in the spectra having different baseline fluxes.  To normalise the individual spectra to a common flux level, every spectrum was divided through its median value.  The median value of a spectrum was used to minimize the influence of cosmic ray hits. This scaling can be performed because this analysis does not depend on the absolute flux, but instead only on the relative changes in flux as a function of wavelength.

\item Alignment of spectra. It is important for our analysis that all of the individual stellar spectra are in the same intrinsic wavelength frame. Since the radial component of the barycentric velocity changes during an observing night, and the absolute wavelength solution of UVES is unstable at the subpixel level, the spectra need to be re-aligned to a common wavelength frame. To do this, Gaussians were fitted to narrow stellar lines close to the sodium D lines in each spectrum to determine the offset relative to a Kurucz model stellar spectrum with atmospheric parameters of $\mathrm{T_{eff}}$ = 5000 K, log(g) = 4.5 \citep{Castelli2004} that was Doppler shifted to account for the system velocity of 55 Cnc of $27.58 \pm 0.07$ $\mathrm{km s^{-1}}$.  These offsets were then used to update the intrinsic wavelength solution for the star. The normalized and aligned spectra are shown in the top panel of Fig. \ref{fig:pcaremoved}.

\item{Removal of cosmic rays. The standard UVES data reduction recipes do not remove cosmic ray hits for observations made with the image slicer.  Therefore, after the UVES spectra were normalized and aligned, cosmic rays were removed by fitting a linear function at each wavelength position through all spectra, so that cosmic rays could be identified as being outliers from the fit.  They were then replaced with the interpolated value from the fit.  This process was iterated twice with different threshold values so it only identified very strong cosmic rays on the first iteration.  This was necessary because the presence of very strong cosmic rays could skew the linear fit and cause weaker cosmic rays to be missed.}

\item{Removal of the stellar absorption features. The stellar spectrum of 55 Cnc was assumed to be constant during a night of observations.  This allowed the stellar features to be removed by dividing every observed spectrum from a single night by the median of all of the observed spectra from that night.  This only slightly weakened the strength of potential planet absorption lines because they changed wavelength significantly ($>$100 pixels) during the transit. The resulting spectra are shown in the second panel of Fig.\ref{fig:pcaremoved}.}

\item{Removal of large systematic trends. Significant systematic trends in the residual spectra in the wavelength direction became apparent after the stellar features had been removed. These trends, which differed for different spectra, were fitted with a linear slope, that was subsequently removed at the beginning of Step 2 (see above). Steps 2 to 5 were redone, after which we proceeded with Step 7.} 

\item{Removal of telluric lines with principal component analysis (PCA). Telluric absorption lines change in strength, mainly due to the change in airmass during observations, but also possibly due to variations in the water vapour content of the  Earth's atmosphere. We removed the telluric absorptions in the sodium D region using PCA (also know as singular value decomposition) over the time domain.   This method relies on the assumption that all of the telluric lines vary in the same way and is discussed in Section 2 of \cite{dekok2013}.  Since Step 3 of our data analysis aligned the spectra to the stellar rest frame, the telluric lines show a small shift in position during the night since they are in the rest frame of the observer. However, the PCA algorithm was able to mostly remove the misaligned telluric lines, as shown in the third panel of Fig. \ref{fig:pcaremoved}, while the components that were removed are shown in Fig. \ref{fig:pca_components}. Some weak residual features from the telluric lines remain after the PCA.  These are probably caused by the line width of the telluric lines changing slightly during the night.  The PCA algorithm is a blind process so if it is allowed to remove a large number of components, it will eventually remove all variation in the data, including the planet signal. However, only four PCA components were required to remove the telluric lines. By injecting artificial planet signals (see Section \ref{subsect:injected} and the lower panel of Fig. \ref{fig:pcaremoved}) we show that the planet signal is left intact by this procedure. 

\begin{figure*}[h] 
\centering 
\includegraphics[width=15cm]{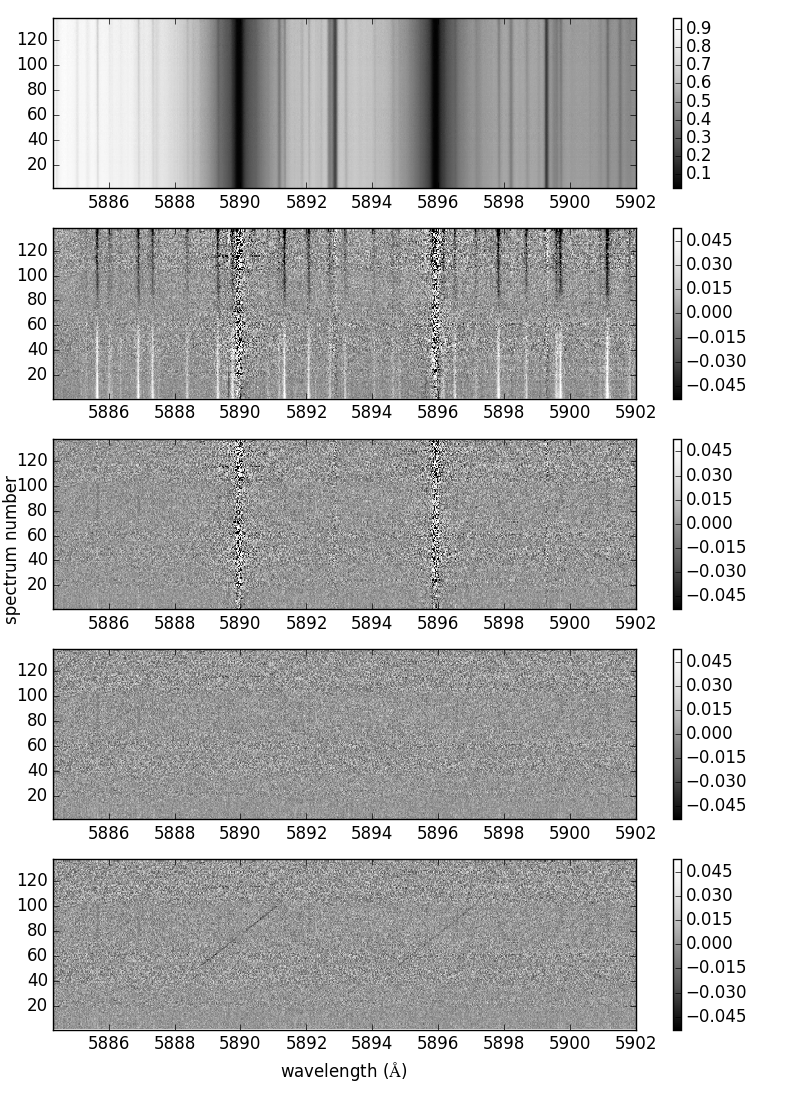}
\caption{Visual representation of the processing steps as described in Section 3. The vertical axis of each matrix represents the sequence number of the observed spectrum. This figure shows the UVES data, but the HARPS and HARPS-N datasets look very similar. The first panel shows the data around the sodium D lines after normalization and alignment in Steps 2 and 3. The second panel shows the residual matrix after dividing through the average star spectrum (Step 5). The third panel shows the residuals after the PCA analysis (Step 7), and the fourth panel shows the same after normalizing each column of the matrix by its standard deviation (Step 8). The bottom panel shows the same data, but after injecting an artificial planet signal before Step 3 - at a level of 3\% of the stellar flux. The injected planet signal can be seen as a diagonal trace from spectrum number 53 to 99, resulting from the change in the radial component of the planet orbital velocity during transit.}
\label{fig:pcaremoved} 
\end{figure*}

\begin{figure*}[h] 
\centering 
\includegraphics[width=15cm]{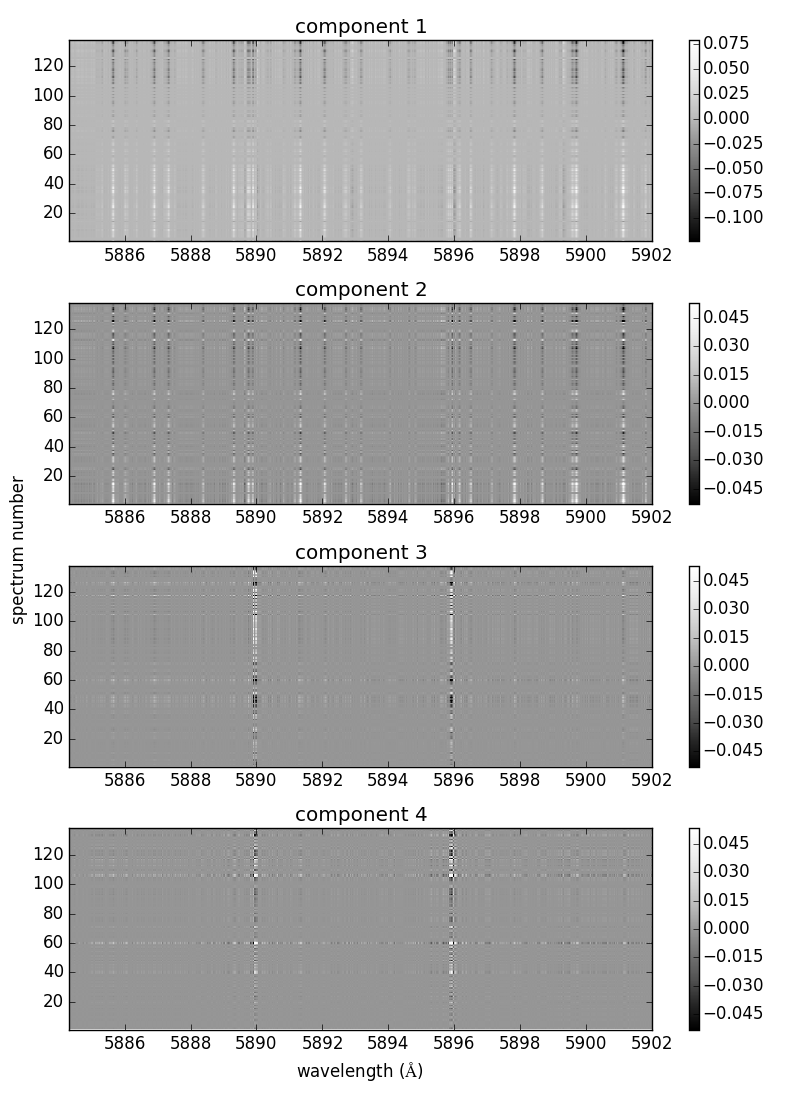}
\caption{Components that were removed by the principle component analysis (PCA) algorithm to remove telluric lines around the Na D lines in the UVES data set.  The top panel shows the first component and subsequent panels show the subsequent components, until the fourth component in the bottom panel. The top panel has a different colour scale to the other panels.}
\label{fig:pca_components} 
\end{figure*}}

\item{Weighting by noise as a function of wavelength. The values at each wavelength position were subsequently weighted down by the noise, derived from the standard deviation of the residuals as function of time at that position. Practically, this only influences the region directly around the cores of the two stellar sodium D lines, as can be seen by comparing panels 3 and 4 of Fig. \ref{fig:pcaremoved}. This naturally weighs down the contribution from the spectra during which the planet absorption overlaps with the strong stellar sodium lines.  The effect this has on the planet signal is illustrated in panel 5 of Fig. \ref{fig:pcaremoved}.  An artificial planet signal is injected into the observed spectra (see section \ref{subsect:injected}) as a 3\% absorption relative to the stellar spectrum.  This signal is weighted with a box-shaped transit profile which is reasonable for a small planet such as 55 Cnc e.  At the mid-transit point when the planet signal is at the same wavelength as the stellar signal (and thus falls in the cores of the Na D lines), the retrieved planet signal is reduced by a factor of approximately} 10.
\end{enumerate}

\subsection{Processing of HARPS and HARPS-N data}

Except for small differences, the processing of the HARPS and HARPS-N data was performed in a similar way to that of the UVES data explained above. Since the data retrieved from the data archives of both telescopes is completely reduced and wavelength-calibrated, Step 1 was not needed. In addition, the wavelength calibration of both HARPS and HARPS-N is stable at the 1 m sec$^{-1}$ level, and delivered to the user in the restframe of the star. Therefore also step 3 was not needed.  

In addition to the sodium D lines (5889.95\AA\  and 5895.92\AA), the HARPS and HARPS-N spectra also cover the Ca H and K lines (at 3968.47\AA\ and 3933.66\AA\ respectively). While the sodium lines are covered by all data sets, the ionized calcium H and K lines are only present in the HARPS and HARPS-N data and not in the UVES data.

\subsection{Combining the different data sets \label{subsection:combining_Na_signal}}

The final step in the analysis is to merge the signal from the two sodium D lines (and calcium H and K lines) and combine the signal from all the in-transit spectra. In addition, we also combine the data sets from the different telescopes.

Two regions of 16$\AA$ centred on the positions of the Na D$_2$ and D$_1$ lines in the residual spectra were averaged with weights proportional to the relative theoretical line strengths. Each spectrum was subsequently shifted to the planet rest frame and added in time over the transit. The signal from the different data sets was subsequently combined using weights proportional to the total in-transit S/N of the data set. The unfavourable observing conditions during the UVES observations caused the UVES observations to have a comparable total S/N to the HARPS and HARPS-N observations (see Table \ref{table:SNR}).

The final data from the calcium H and K lines were produced in the same way using a weighting ratio of Ca K/Ca H = 2 for the two lines.

\begin{table}[h]
\tiny
\centering
\caption{Signal-to-noise-ratios (S/N) of the data sets.} 

\label{table:SNR}

\begin{tabular}{lccc}
\hline \hline 
\pbox{20cm}{dataset} & \pbox{20cm}{average S/N \\ per spectrum} & \pbox{20cm}{number of spectra \\ during transit} & \pbox{20cm}{total S/N} \\ \hline 
UVES  & 116.3 & 47 & 977  \\
HARPS-N B & 161.2 & 18  & 838 \\
HARPS-N A & 223.3 & 19 & 1192 \\
HARPS B & 109.6 & 24 & 658 \\
HARPS A & 143.9 & 24 & 863 \\
\hline
\end{tabular}

\end{table}

\subsection{Injection of artificial planet signals \label{subsect:injected}}

A useful technique to determine the magnitude of the absorption signal of the planet relative to the stellar spectrum is to  inject artificial planet signals early on in the data processing so that the artificial signals are processed in the same way as a real signal would be.  This also allows us to check to what level our analysis removes any planet signal.  The injection of artificial signals was carried out in a similar way to \cite{Snellen2010} and \cite{Hoeijmakers2015}.

The artificial planet signals of the sodium $\mathrm{D}_1$ (5895.92 \AA) and $\mathrm{D}_2$ (5889.95 \AA) lines were generated using Gaussian line profiles of equal width, and with amplitudes with a ratio of $\mathrm{D}_2/\mathrm{D}_1 = 2$.  These relative line strengths were calculated according to equation 1 in \cite{Sharp2007}. We do note that these equations in principle only hold for local thermodynamic equilibrium, while planet exospheres may be better described by radiative transfer algorithms which use a non-Maxwellian velocity distribution function such as in \cite{Chaufray2013}. The quantum parameters which describe the Na D line transitions were obtained from the Vienna Atomic Line Database (VALD) \citep{Kupka2000}. The relative strengths of the Na D lines are practically independent of temperature across the range of 1000 to 3000 K. We assumed $\mathrm{T} = 2000$ $\mathrm{K}$. 
 
Using the orbital parameters from \cite{Gillon2012} and assuming a circular orbit \citep{Demory2012}, the radial velocity of the planet can be calculated at the time of each exposure to determine the central wavelengths of the Doppler-shifted sodium lines. 

The planet signal was injected into the in-transit data by multiplying the observed spectra with the artificial absorption model according to

\begin{equation}
\rm F(\lambda)_{\rm injected} = {[1 - A \times F_{\rm model}(\lambda,\rm v_{rad})]F_{\rm obs}(\lambda)},
\end{equation}

where $\rm F_{obs}(\lambda)$ is the observed spectrum, $\rm F_{model}(\lambda,v_{rad})$ is the Doppler shifted sodium model spectrum, with $\rm A$ as a scaling parameter that sets the amplitude of the sodium D$_2$ line, and $\rm F(\lambda)_{injected}$ is the output spectrum.

During an exposure, the radial component of the orbital velocity of the planet changes significantly. For example, the observations taken with HARPS-N have an exposure time of 240 seconds, during which the planet radial velocity changes by approximately 5 km sec$^{-1}$, corresponding to six resolution elements. Thus, even for an intrinsically narrow planet absorption, the observed signal cannot be narrower than five or six pixels. Therefore, the injected artificial sodium lines have a width equal to this observational broadening, which is different for each data set, as shown in Table \ref{tab:widths}.

The same procedure was used to inject an artificial absorption signal of ionized calcium, using a relative line ratio of Ca K/Ca H = 2 as calculated from \cite{Sharp2007}.

\begin{table}[]
\centering
\caption{An estimation of the width of the absorption signal from 55 Cnc e based on the average change of its radial velocity during the exposure of each spectrum in each data set.}
\label{tab:widths}
\begin{tabular}{lccc}
\hline\hline 
Dataset & width $\mathrm{(kms^{-1})}$ & width (\AA) & width (pixels)\\ \hline
UVES & 1.3 & 0.026 & 1.0 \\
HARPS A &  4.0  & 0.078 & 4.9\\
HARPS B & 4.0 & 0.078 &  4.9\\
HARPS-N A & 5.3 & 0.105 & 6.5 \\
HARPS-N B & 5.3 & 0.105 & 6.5\\\hline
\end{tabular}
\end{table}

\begin{table*}[h]
\tiny
\centering
\caption{Observational timing parameters.  The orbital phases of 55 Cnc e are based on the orbital parameters derived from the SPITZER + MOST observations used in \cite{Gillon2012}.}
\label{table:timing}

%\begin{tabular}{@{}c@{}||@{}c@{}|@{}c@{}|@{}c@{}|@{}c@{}|@{}c@{}}
\begin{tabular}{l l l l l l l}
\hline \hline 
\pbox{20cm}{data set} & \pbox{20cm}{UVES} & \pbox{20cm}{HARPS-N B} & \pbox{20cm}{HARPS-N A} & \pbox{20cm}{HARPS B} & \pbox{20cm}{HARPS A} \\

\pbox{20cm}{Program nr.} & \pbox{20cm}{ESO: 092.C-0178} & \pbox{20cm}{TNG: CAT13B\_33} & \pbox{20cm}{TNG: CAT13B\_33} & \pbox{20cm}{ESO: 288.C-5010} & \pbox{20cm}{ESO: 288.C-5010} \\

\hline 

\pbox{20cm}{date (UTC)} & \pbox{20cm}{2014-01-04} & \pbox{20cm}{2013-11-29} & \pbox{20cm}{ 2013-11-15} & \pbox{20cm}{2012-03-16} & \pbox{20cm}{2012-01-28} \\

\pbox{20cm}{start \\ phase}  & 0.871 &  0.850 & 0.850 &  0.944 & 0.939 \\  
\pbox{20cm}{end \\ phase}  & 0.106 & 0.108 & 0.074 & 0.077 & 0.093 \\  
\pbox{20cm}{cadence (s)}  & 109.0 & 265.8  & 264.3 & 211.4 & 211.4  \\ 
\pbox{20cm}{exposure \\ time (s)} & 60 & 240 & 240 & 180 & 180$^\dagger$ \\ \hline
\pbox{20cm}{observation \\ start (UTC)} & 04:43:03.805 & 02:10:18.576 & 02:18:54.502 & 01:01:25.593 & 03:56:16.478 \\  
\pbox{20cm}{transit \\ start (UTC)} & 06:16:30.913 & 04:05:54.424 & 04:14:02.316 & 01:17:26.886 & 04:16:50.727 \\
\pbox{20cm}{mid-transit \\ time (UTC)} & 07:00:13.153 & 04:49:36.664 & 04:57:44.556 & 02:01:09.126 & 05:00:32.967 \\ 
\pbox{20cm}{transit \\ end (UTC)} & 07:43:55.393 & 05:33:18.904 & 05:41:26.797 & 02:44:51.366 & 05:44:15.207 \\

\pbox{20cm}{observation \\ end (UTC)} & 08:52:37.259 & 06:44:16.990 & 06:15:53.853 & 03:22:45.983 & 06:39:14.017  \\ \hline
\pbox{20cm}{Nr. spectra \\ pre-transit} & 52 & 26 & 25 & 5 & 6 \\ 
\pbox{20cm}{Nr. spectra \\ in transit} & 47 & 18 & 19 & 24 & 24 \\ 
\pbox{20cm}{Nr. spectra \\ post-transit} & 39 & 17 & 9 & 12 & 17 \\ 
\pbox{20cm}{total nr. \\ of spectra} & 138 & 61 & 53 & 41 & 47 \\ 
\hline
\multicolumn{6}{l}{$\dagger$ Except for the first two spectra which have exposure times of 123s}
\end{tabular}

\end{table*}

\section{Results \label{sec:results}}

\subsection{Sodium}

The results for sodium are shown in Fig. \ref{fig:all_datasets_no_injected}. The left and right panels show the unbinned data and data binned by five pixels (or 0.05 $\AA$) respectively. From top to bottom the panels show the two HARPS data sets, the two from HARPS-N, the UVES data set, and the signal combined from all telescopes.  In these panels, the stellar and telluric signals have been removed so all that remains is residual noise and a possible absorption signal from 55 Cnc e.   The noise has an approximately Gaussian distribution, so the statistical significance of the detection can be estimated by comparing the depth of the absorption signal to the standard deviation of the noise.  Since any planet signal is expected to be broadened due to the long exposure times, the S/N in the unbinned data, calculated as described above, may be underestimated. 

While there is a hint of planet absorption in the individual UVES data set, this is somewhat more pronounced in the combined (binned) data.   This signal has a statistical significance of 3.2$\sigma$ and $3.3\sigma$ in the unbinned and binned data respectively. The binned and unbinned versions of the combined data are also overlayed in Fig. \ref{fig:binned7perbinNoInjected} for clarity. 

By injecting artificial signals at various levels relative to the stellar spectrum we can estimate the strength of the retrieved signal. If real, the planet sodium lines in the combined data are at a level of 2.3$\times 10^{-3}$ with respect to the star. 

\subsection{Ionized calcium}

\begin{figure*}
\centering 
\includegraphics[width=15cm]{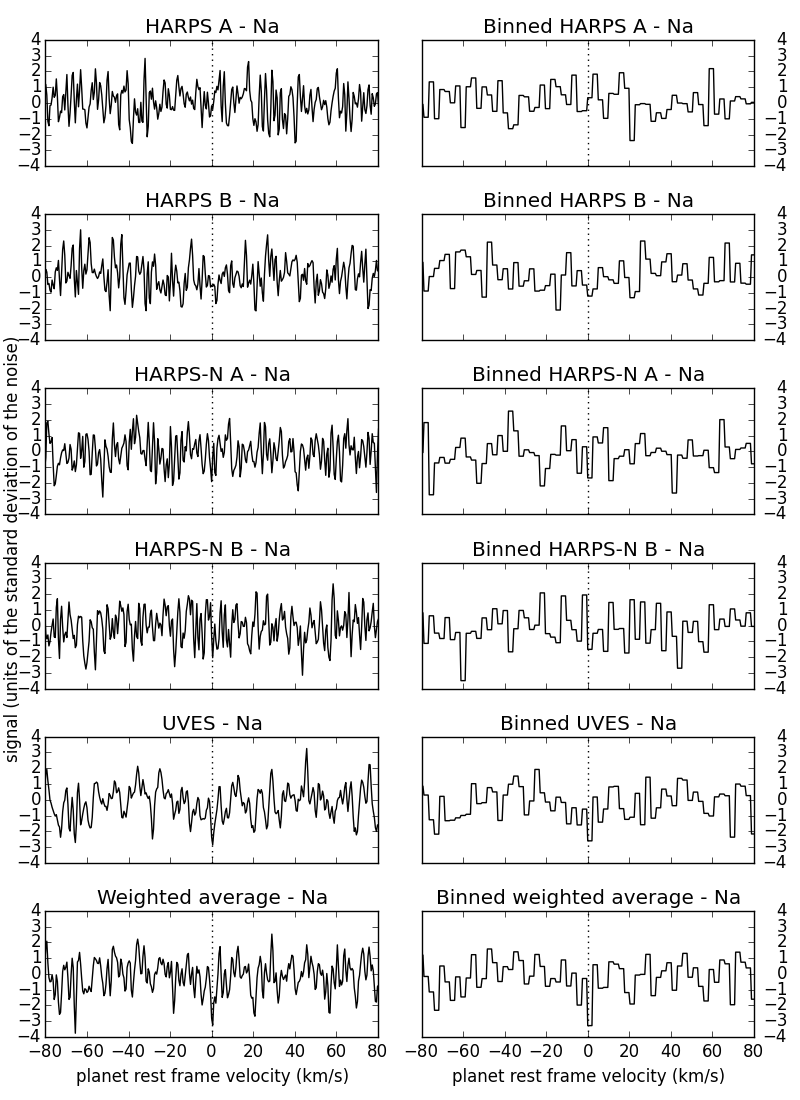} 
\caption{Signal of sodium from the individual data sets and from an average of the data sets, weighted  according to the S/N of the data set.  The signals have been scaled to be in units of the standard deviation of the noise.  The left column is not binned while the right column is binned every 0.05\AA\ or 3.8 km sec$^{-1}$.  The binned average signal has a detection S/N = 3.3.  The vertical dotted line in all panels indicates a planet rest frame velocity of 0 km sec$^{-1}$.}
\label{fig:all_datasets_no_injected} 
\end{figure*}

\begin{figure*}[h] 
\centering 
\includegraphics[width=12cm]{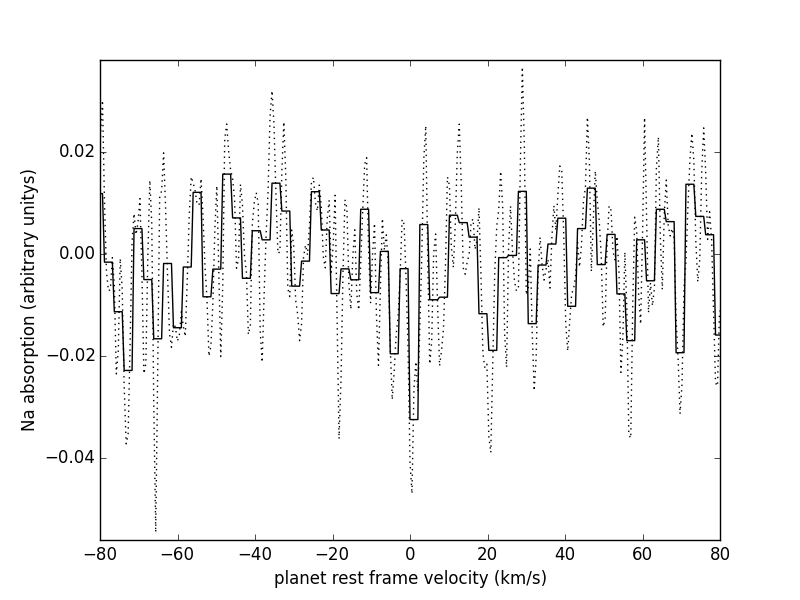} 
\caption{Average signal of sodium from all data sets both not binned (dotted line) and binned (solid black line) every 0.05\AA\ or 3.8 km sec$^{-1}$ (as shown in the bottom panels of Fig. \ref{fig:all_datasets_no_injected}).  This binning regime results in a detection that has a S/N of approximately $3.3\sigma$.}
\label{fig:binned7perbinNoInjected}   
\end{figure*}

\begin{figure*}
\centering 
\includegraphics[width=15cm]{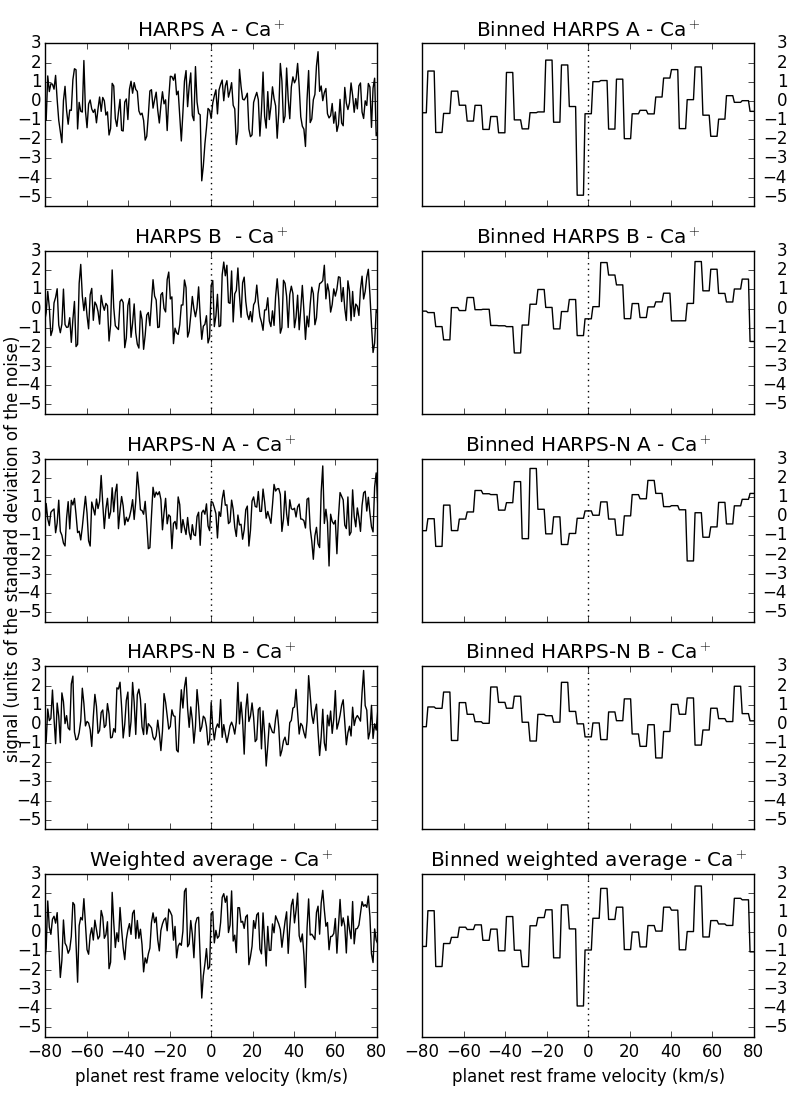} 
\caption{Signal of ionized calcium from the individual data sets and from an average of the data sets, weighted  according to the S/N of the data set.  The signals have been scaled to be in units of the standard deviation of the noise.  The left column is not binned while the right column is binned every 0.05\AA\ or 3.8 km sec$^{-1}$.  The vertical dotted line in all panels indicates a planet rest frame velocity of 0 km sec$^{-1}$.  The binned average signal has a detection S/N = 4.1; however, this completely originates from the HARPS A data set which individually has a binned detection S/N of 4.9.}
\label{fig:CaIIAllDatasets} 
\end{figure*}

\begin{figure*}[h] 
\centering 
\includegraphics[width=12cm]{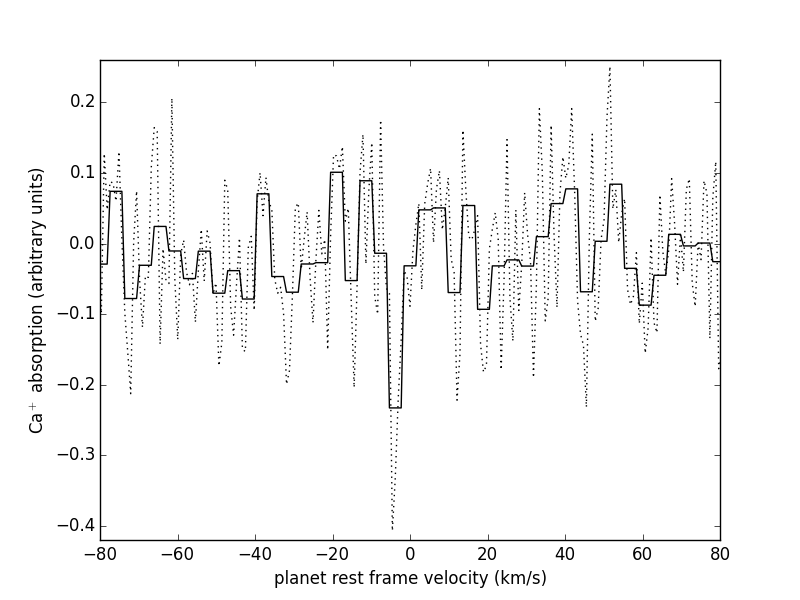} 
\caption{Signal of ionized calcium from the HARPS A data set both not binned (dotted line) and binned (solid black line) every 0.05\AA\ or 3.8 km sec$^{-1}$ (as shown in the bottom panels of Fig. \ref{fig:CaIIAllDatasets}).  This binning regime results in a detection that has a S/N of approximately 4.9$\sigma$.}
\label{fig:CaOverlay} 
\end{figure*}

The results for ionized calcium are shown in Fig. \ref{fig:CaIIAllDatasets}. The individual panels are the same as in Fig. \ref{fig:all_datasets_no_injected}, except that UVES is not included because the wavelength range of the UVES data does not cover the calcium H and K lines. In contrast to the sodium data, an interesting signal can be seen in the first HARPS-A data set. It shows a feature that has a statistical significance of 4.9$\sigma$, although it is blueshifted with respect to the planet rest frame by approximately 4 km sec$^{-1}$. An overlay of the binned and unbinned data of the HARPS-A data set is shown in Fig. \ref{fig:CaOverlay}. The signal does not appear in the other data sets, resulting in a S/N of less than 4 in the combined data. 

\begin{figure*}[h] 
\centering 
\includegraphics[width=15cm]{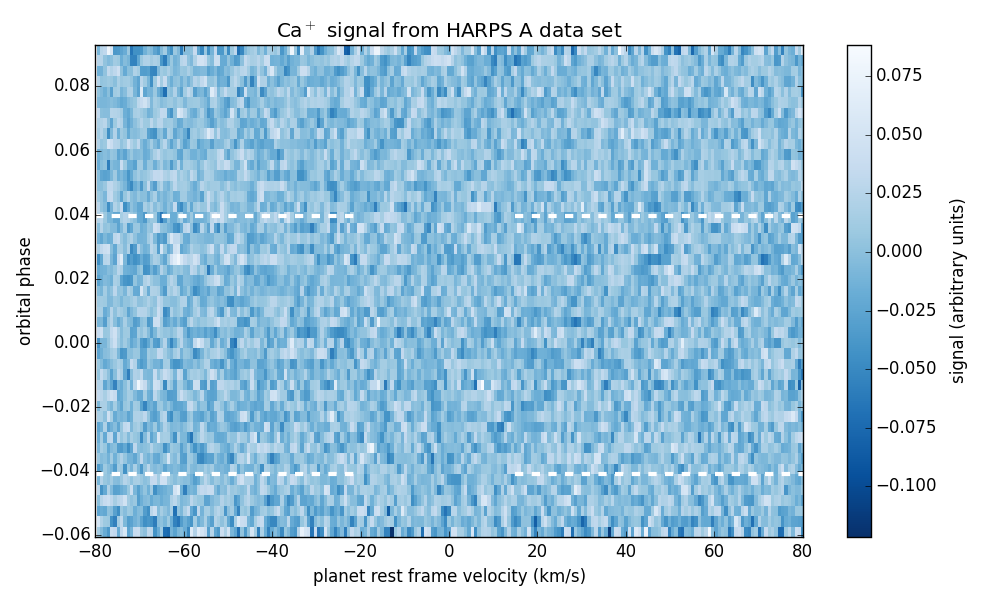} 
\caption{Trace of the signal of Ca$^+$ from the HARPS A data set across the time-series of spectra in the rest frame of 55 Cnc e.  In this frame, the planet signal lies on a vertical line, blueshifted by approximately 4 km sec$^{-1}$.  Dashed horizontal white lines indicate the transit duration.}
\label{fig:Camatrix} 
\end{figure*}

The contribution to the 4.9$\sigma$ Ca$^+$ signal from each individual spectrum of the HARPS-A data set is shown in Fig. \ref{fig:Camatrix}.  This figure presents the data in the rest-frame of 55 Cnc e so that the features that contribute to the signal lie on a vertical line that is blueshifted by approximately 4 km sec$^{-1}$. The transit duration is indicated in the figure.  
It can be seen that there are contributions from multiple spectra during transit, indicating that the signal is not caused by a random spurious feature in a single spectrum.  If the exosphere is extended beyond the Roche lobe, one would expect it to be distorted and hence possess different velocities relative to the planet and possibly be detectable just before or after transit.  However, the S/N in the data is not sufficient to see such distortions or extended absorption signatures.

If real, the planet calcium H and K lines in the HARPS-A data set are at a level of $7.0\times 10^{-2}$ with respect to the star. 

\begin{figure*}
\centering 
\includegraphics[width=15cm]{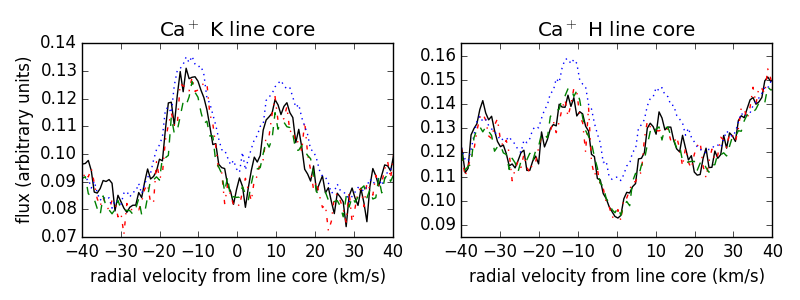} 
\caption{Median spectrum of each data set showing the median emission from the cores of the Ca$^+$ H (right) and K (left) lines.  The solid black line is the HARPS A data set, the dotted-dashed red line is the HARPS B data set, the dotted blue line is the HARPS-N A data set and the dashed green line is the HARPS-N B data set. The two panels have different vertical scales.}
\label{fig:CaHandK_cores} 
\end{figure*}

To assess whether the Ca$^+$ signal could originate from variability in the stellar Ca$^+$ H and K lines, we investigated the emission from the cores of the H and K lines in all data sets (Fig. \ref{fig:CaHandK_cores}). We found no evidence for variations in excess of that expected from Poisson noise within each transit data set. Although one dataset (HARPS-N A, hence not corresponding to that showing Ca$^+$) exhibits stronger Ca H+K emission (55 Cnc has a known 39 day periodicity in its Ca$^+$ H and K stellar emission lines; \cite{Fischer2008}), it shows no variability during the transit. Also, the radial velocity of 55 Cnc e changes by $\pm$57 km sec$^{-1}$ which causes a Doppler shift of $\pm$0.75\AA\ relative to the core of the lines.  Since the signals across spectra are combined in the planet rest frame, only the spectra taken close to the mid-transit point (where the planet signal is at the same wavelength as the stellar lines) could be influenced by variability in the Ca$^+$ H and K emission.  Therefore, even if there was some variability in emission during a night of observations, its impact on the results would still be limited.

\section{Discussion and conclusions \label{sec:discussion_and_conclusion}}
In this paper we carried out a search for neutral sodium (Na) and singly ionized calcium (Ca$^+$) in the exosphere of the exoplanet 55 Cnc e with transmission spectroscopy.  This search yielded a $3.3\sigma$ detection of Na after combining five transit data sets and a 4.9$\sigma$ detection of Ca$^+$ in only one transit data set.

We estimated the p-value of the Ca$^+$ detection in one of our four HARPS(-N) data sets. The probability of observing a spurious 4.9 $\sigma$ signal is very low at 4.8$\times 10^{-7}$. However, we would have observed such a signal at any velocity between $-$50 and +50 km sec$^{-1}$ in the planet rest frame, corresponding to about approximately 20 positions. Multiplying this by the number of transits observed means that we had approximately 80 opportunities to observe a spurious signal, meaning that we can estimate that the false alarm probability is $\sim80 \times 4.8\times 10^{-7} \approx 4 \times 10^{-5}$, corresponding to $<4 \sigma$. For this estimate we do not take into account our freedom to use a certain width for the probed signal. In addition, the possible impact of unquantified correlated noise in the data may also increase the p-values. We therefore think these data are as yet insufficient to claim definite detections of the planet exosphere. 

As discussed above, the spectral resolution of any potential planet signal is broadened due to the change of the radial component of the planet orbital velocity during an exposure. As shown in Table \ref{tab:widths}, this `instrumental' broadening is five to six pixels for the HARPS and HARPS-N data, and below the intrinsic spectral resolution of the spectrograph for the UVES data, due to the significantly shorter exposure times. In addition, the absorption from sodium and ionized calcium could be intrinsically broadened due to a strong velocity field in the planet exosphere \citep{Mura2011}.   If the observed signals from either sodium or ionized calcium are real, they are indeed broad at the five to six pixel (4 km sec$^{-1}$) level, which is much broader than the intrinsic width of the Na D$_2$ line in the exosphere of Mercury, previously observed by \cite{Potter2013}, of approximately 20 m$\AA$ or approximately 1 km sec$^{-1}$.

It is not clear whether to expect the blueshift of approximately 4 km sec$^{-1}$ as measured for the potential ionized calcium signal. On the one hand, the photon-ionization lifetime of singly-ionized calcium is estimated to be significantly longer than that of neutral sodium,  which could allow it to accumulate a significant acceleration as it is picked up by the stellar wind and dragged in the anti-stellar direction. On the other hand, if the planet has a significant dipole magnetic field, the ionized calcium may be trapped in the planet's magnetic field \citep{Mura2011}.  

If the ionized calcium signal is real, it would imply that this signal is highly variable, since it is only visible in one of the four data sets. As discussed in the introduction, we do know that Mercury's exosphere is highly variable, on a range of timescales from days, to months, to years \citep{Killen2007}. It is not clear at this stage whether we would expect similar behaviour for the exospheres of hot rocky super-Earths.  If 55 Cnc e were to have a significant atmosphere, a confirmed detection would be evidence of atmospheric blow-off.  On the other hand, if 55 Cnc e does not have a thick atmosphere, as is suggested by \cite{Demory2016} as being a likely interpretation of its longitudinal thermal brightness map, the exosphere would likely be produced by sputtering of the surface.

We modelled the exosphere of 55 Cnc e to first order as an optically thick ring around the planet. Ignoring subtle effects like stellar limb darkening, the fraction of starlight absorbed by sodium and ionized calcium in the exosphere of 55 Cnc e would correspond to an outer radius of the exosphere of 5 $\mathrm{R_\oplus}$ and 25 $\mathrm{R_\oplus}$ respectively - 2.3 and 12 times the radius of the planet. We compare this to the Roche radius of 55 Cnc e, calculated using 

\begin{equation}
R_R = \frac{0.49q^{2/3}}{0.6q^{2/3} + \mathrm{ln}(1+q^{1/3})},
\end{equation}

where $q = M_{planet}/M_{star}$ \citep{Eggleton1983}, which is found to be $R_R = 5.35$ $R_\oplus$. Hence, the possible sodium signal, if optically thick, would come from a region as large as the planet's Roche lobe, while that of ionized calcium would be significantly larger. If the sodium exosphere were not optically thick, it would also need to be significantly larger than the planet's Roche lobe.  If the Ca$^+$ exosphere really were to have a radius of 25 $\mathrm{R_\oplus}$, it would have an earlier ingress and a delayed egress compared to what would be expected from the radius of the planet as determined by broadband photometry.  Using \cite{Mandel2002}, the transit duration for 55 Cnc e with a 25 $\mathrm{R_\oplus}$ exosphere was found to last 26 minutes longer than the broadband transit duration. This corresponds to a range of orbital phases of -0.054 to 0.054.  There may be a  hint of this early ingress and delayed egress, as can be seen in Fig. \ref{fig:Camatrix}, however, the S/N is not sufficient to make definite claims.     

 Although no formal detection has been made, we advocate that probing the exospheres of hot super-Earths in this way has great potential, also knowing that Mercury's exosphere varies significantly over time. It may be a fast route towards the first characterization of the surface properties of this enigmatic class of planets. Our team is pursuing a transit monitoring programme with UVES to further investigate the possible variable signal from ionized calcium. 

%%%%%%%%%%%%%%%%%%%%%%%%%%%%%%%%%%%%%%%%%%%%%%%%%%%%%%%%%%%%%%%%%%%%%%%%%%%%%%%%%%%%%%%%%%%%%%%%%%%%

%-------------------------------------------------------------------
\bibliographystyle{aa} % style aa.bst

\begin{acknowledgements} 
A. R. R.-H. is grateful to the Planetary and Exoplanetary Science (PEPSci) programme of the Netherlands Organisation for Scientific Research (NWO) for support.  I. A. G. S. acknowledges support from an NWO VICI grant (639.043.107). M. B. acknowledges support by NASA, through Hubble Fellowship grant HST-HF2-51336 awarded
by the Space Telescope Science Institute.  We thank the anonymous referee for their constructive comments. 
\end{acknowledgements}

\bibliography{55Cnce}

\end{document}